\documentclass[11pt]{article}
\usepackage{amsmath,amssymb}

\begin{document}
\tolerance=5000
\newcommand{\s}{\hspace{0.15cm}}
\newcommand{\bc}{\bigcirc}
\newcommand{\ba}{\begin{alignat}{3}}
\newcommand{\al}{\alpha}
\newcommand{\bt}{\beta}
\newcommand{\gm}{\gamma}
\newcommand{\dl}{\delta}
\newcommand{\ep}{\epsilon}
\newcommand{\sg}{\sigma}
\newcommand{\om}{\omega}
\newcommand{\pa}{\partial}
\newcommand{\tf}{\tfrac}
\newcommand{\mc}{\mathcal}
\newcommand{\op}{\oplus}
\newcommand{\ol}{\overline}
\newcommand{\we}{\wedge}

\begin{titlepage}
\begin{flushright}
    OUHET--578 \\
    hep-th/0703154  \\
    March 2007
\end{flushright}
\begin{center}
  \vspace{3cm}
  {\bf \Large Toward the Determination of $R^3F^2$ Terms in M-theory}
  \\  \vspace{2cm}
  Yoshifumi Hyakutake\footnote{E-mail: hyaku@het.phys.sci.osaka-u.ac.jp}
   \\ \vspace{1cm}
   {\it Department of Physics, Graduate School of Science, Osaka University \\
   Machikaneyama 1-1, Toyonaka, Osaka 560-0043, Japan}
\end{center}

\vspace{2cm}
\begin{abstract}
  Higher derivative terms in M-theory are investigated by applying
the Noether method. Cancellation of variations under the local supersymmetry
is examined to the order linear in $F$ by a computer program. Structure of $R^4$ terms
is uniquely determined and exactly matches with one-loop effective terms
in type IIA superstring theory. A part of $R^3F^2$ terms is also determined.
\end{abstract}
\end{titlepage}

\setlength{\baselineskip}{0.65cm}

\section{Introduction}

Superstring theory is promising as a theory of quantum gravity, and
interactions of strings are described by supergravity in the low energy limit\cite{Y,SS}.
Scattering amplitudes of strings also contain higher derivative interactions which are not 
included in the supergravity\cite{SS}. Since these terms correspond to quantum or stringy effects
of the gravity, it is important to determine the structure of these corrections completely.

Among the higher derivative terms, quadratic and quartic terms of Riemann tensors 
in heterotic string theories and quartic terms of Riemann tensors in type II superstring 
theories are considerably investigated in the last twenty years from various viewpoints.
In heterotic superstring theories, the existence of the $R^2$ terms is crucial to supersymmetrize
the Lorentz Chern-Simons term\cite{RW}-\cite{BR2}. The cancellation of the gravitational anomaly 
inevitably introduce the higher derivative terms $BR^4$, known as Green-Schwarz anomaly 
cancellation terms\cite{GS}, and the $R^4$ terms are necessary to supersymmetrize 
it\cite{RSW1}-\cite{Su}.
The structure of the $R^4$ terms in heterotic string theories are also investigated by 
evaluating tree or one-loop amplitudes in refs.~\cite{CN}-\cite{EJM}.

In type II superstring theories, tree level effective action, 
$e^{-2\phi} (t_8 t_8 R^4 + \tf{1}{8}\epsilon_{10}\epsilon_{10} R^4)$, is obtained by
combining results of four graviton scattering amplitude and 4-loop 
computation in sigma-model\cite{GW,GS2,GVZ}. The effective action at one-loop order consists of
$t_8 t_8 R^4 \mp \tf{1}{8}\epsilon_{10}\epsilon_{10} R^4 - \tf{1}{6}t_8\ep_{10}BR^4$,
where the minus is for type IIA and the plus sign is for type IIB.
The sign is sensitive to the chirality of the theory\cite{AFMN,KP}, and
the last term in the one-loop effective action is introduced to ensure the string-string 
duality between type IIA on K3 and heterotic string on $T^4$\cite{VW,DLM}.
Under this duality, the last term is related to the Green-Schwarz anomaly cancellation term
in the effective action of heterotic superstring theories. 
The local supersymmetry also relates the $BR^4$ terms and the $R^4$ terms\cite{PVW}.
Pure-spinor formalism is also powerful to derive the higher
derivative terms\cite{PT,BM}.
More detailed information and other techniques for searching the higher derivative 
corrections in string theory will be found, for example, in refs.~\cite{PVW,Ts}.

The higher derivative corrections in M-theory are obtained by lifting the 
above type IIA result to eleven dimensions. 
It is also possible to directly confirm the existence of higher derivative terms
in eleven dimensions from the scattering amplitudes of 
superparticles\cite{GV}-\cite{AGV} or superspace formalism\cite{CGNN}-\cite{Ra}.

This paper is a successive work of refs. \cite{HO1,HO2}, where the structure of higher derivative 
terms in M-theory is investigated by applying the Noether method. There the cancellation is 
checked by neglecting variations which depend on 4-form field strength $F$.
In this paper we examine the cancellation of terms which are linearly dependent on $F$.
Since the number of terms in the ansatz and that of the variations are enormous,
a computer program will be employed to complete this task. 
We will see that the local supersymmetry is powerful enough to 
determine the structure of $R^4$ terms uniquely. A part of $R^3F^2$ terms is also determined,
but the cancellation of $\mc{O}(F^2)$ terms will be necessary to fix completely.

The contents of our paper is as follows. In section 2 we briefly review 
$\mathcal{N}=1$, $D=11$ supergravity and $R^4$ corrections from the viewpoint of
local supersymmetry.
In section 3 the ansatz for the higher derivative corrections $R^3F^2$ and
their variations are explained.
In section 4 we show the results obtained by using the computer program.
We find that the structure of $R^4$ terms is uniquely determined and is consistent with 
one-loop scattering amplitudes in type IIA superstring theory. 
Section 5 is devoted to conclusion and discussion.

\section{Review of Eleven Dimensional Supergravity and $R^4$ Corrections}

In this section we briefly fix notations of the eleven dimensional supergravity\cite{CJS},
and review $R^4$ corrections in M-theory\cite{HO1,HO2}.
The supermultiplet of the eleven dimensional supergravity consists of
a vielbein $e^a{}_\mu$, a Majorana gravitino $\psi_\mu$ and a 3-form field $A$, 
and up to the order of $\mc{O}(\psi^4)$ the action is given by
\ba
  2 \kappa_{11}^2 S = \int d^{11}x \; e \Big( R 
  - \frac{1}{2} \bar{\psi}_\rho \gm^{\rho\mu\nu} \psi_{\mu\nu}
  - \frac{1}{2 \cdot 4!} F_{\mu\nu\rho\sg}F^{\mu\nu\rho\sg} \Big)
  - \frac{1}{3!} \int A \we F \we F + \mc{O}(\psi^4). \label{eq:sugra}
\end{alignat}
In this note we use $D_\mu$ for the ordinary covariant derivative which acts on the 
local Lorentz indices, and $\mc{D}_\mu$ for a covariant derivative modified by the 
4-form field strength like
\ba
  \mc{D}_\mu \psi_\nu = D_\mu \psi_\nu + F_\mu \psi_\nu, \qquad 
  F_\mu = - \frac{1}{36} F_{\mu ijk}\gm^{ijk} + \frac{1}{288} F_{ijkl}\gm_\mu{}^{ijkl}.
\end{alignat}
The field strength of the Majorana gravitino is defined by the modified covariant derivative
as $\psi_{\mu\nu} = 2 \mc{D}_{[\mu} \psi_{\nu]}$. The symbol $[\mu_1 \cdots \mu_n]$ 
means that the indices inside the brackets are completely antisymmetrized with a factor
$\frac{1}{n!}$ for each term.
Note that $\mu, \nu$ etc. are space-time indices and $a,b,i,j$ etc. are local Lorentz indices.

The action (\ref{eq:sugra}) possesses $\mc{N}=1$ local supersymmetry in eleven dimensions, 
under which the fields transform as
\ba
  &\dl e^a{}_{\mu} = \bar{\ep} \gm^a \psi_\mu, \qquad
  \dl \psi_\mu = 2 \mc{D}_\mu \ep, \qquad
  \dl A_{\mu\nu\rho} = -3 \bar{\ep} \gm_{[\mu\nu} \psi_{\rho]}. \label{eq:susytr}
\end{alignat}
Here $\ep$ is a space-time dependent parameter which transforms as a Majorana spinor.
In order to check the invariance of the action, it is worth noting the commutation relation
\ba
  [\mc{D}_\mu, \mc{D}_\nu] &= \tf{1}{2} R_{ab\mu\nu} T^{ab} 
  - \tf{1}{18} (D_{[\mu} F_{\nu] ijk}) \gm^{ijk} 
  + \tf{1}{144} (D_{[\mu} F^{ijkl}) \gm_{\nu] ijkl} \notag
  \\
  &\quad\, +
  \tf{1}{1728}F_{ijkl}^2\gm _{\mu\nu} 
  + \tf{1}{216}F_{ijkl}F_{ijk[\mu}\gm_{\nu]l} 
  -\tf{1}{48}F_{\mu ikl}F_{\nu jkl} \gm_{ij} 
  +\tf{1}{72}F_{ijkl}F_{ijm[\mu} \gm_{\nu]klm} \notag
  \\
  &\quad\,
  + \tf{1}{108} F_{\mu\nu ij} F_{iklm} \gm_{jklm} 
  -\tf{1}{576} F_{ijkl}F_{ijmn} \gm_{\mu\nu klmn} 
  -\tf{1}{432}F_{ijkl}F_{imn[\mu} \gm_{\nu]jklmn} \label{eq:comm}
  \\
  &\quad\,
  +\tf{1}{864}F_{\mu ijk}F_{\nu lmn} \gm_{ijklmn} 
  -\tf{1}{2592}F_{ijkl}F_{mno[\mu}\gm_{\nu]ijklmno} 
  +\tf{1}{41472}F_{ijkl}F_{pqrs} \gm_{\mu\nu ijklpqrs}, \notag
\end{alignat}
where $T^{ab}$ is a generator of the Lorentz algebra,
$(T^{ab})_{ij} = 2 \dl^{[a}_i \dl^{b]}_j$ for vector representation
and $T^{ab} = \tf{1}{2} \gm^{ab}$ for spinor one.
The variation of the kinetic term for the Majorana gravitino is equal
to $- 2\bar{\psi}_a \gm^{abc} [\mc{D}_b, \mc{D}_c] \ep$ up to $\mc{O}(\psi^2)$,
and after some calculations it becomes
\ba
  - 2\bar{\psi}_a \gm^{abc} [\mc{D}_b, \mc{D}_c] \ep &= E(e)^{ab} \bar{\ep} \gm_b \psi_a 
  + 3 E(A)^{abc} \bar{\ep} \gm_{bc} \psi_a + \mc{O}(\psi^2). \label{eq:eqmajo}
\end{alignat}
Here $E(e)_{ab}$ and $E(A)^{abc}$ are field equations of the vielbein and the 3-form field. 
Explicit expressions of the field equations are written as
\ba
  E(e)_{ab} &= 
  2 R_{ab} - \eta_{ab} R - \tf{1}{6} F_{aijk} F_{bijk} 
  + \tfrac{1}{48} \eta_{ab} F_{ijkl} F^{ijkl}, \notag
  \\
  E(\psi)^a &= - \gamma^{abc} \psi_{bc}, \label{eq:EoM}
  \\
  E(A)^{abc} &= \tf{1}{6} D_d F^{dabc} 
  - \tf{1}{48 \cdot 144} \ep_{11}^{abcijklmnop} F_{ijkl} F_{mnop}. \notag
\end{alignat}
$E(\psi)^{a}$ is the field equation of the Majorana gravitino.
Thus the right hand side of (\ref{eq:eqmajo}) can be cancelled by the variations of the bosonic 
terms and the invariance under the local supersymmetry is shown up to $\mc{O}(\psi^2)$.
Note also that the integrability condition $[\mc{D}_\mu, \mc{D}_\nu]\ep=0$ automatically 
satisfies the classical equations of motion. 

Now let us review the higher derivative corrections to the eleven dimensional supergravity
from the viewpoint of the local supersymmetry. 
Corrections start from terms with mass dimension eight which are composed of 
the Riemann tensors, the 4-form field strengths, the covariant derivatives and 
the Majorana gravitinos.
The mass dimensions of these components are 2, 1, 1 and $\frac{1}{2}$, respectively.
The duality between type IIA superstring theory and M-theory suggests that the ansatz for
the higher derivative terms should include the $R^4$ terms.
Furthermore the ansatz will be simplified by using the field redefinition ambiguity.
That is, with the aid of the relations
\ba
  R &= -\tf{1}{9} E(e)^a{}_a + \tf{1}{144} F_{ijkl}F^{ijkl}, \notag
  \\
  R_{ab} &= \tf{1}{2} E(e)_{ab} - \tf{1}{18} \eta_{ab} E(e)^i{}_i 
  + \tf{1}{12} F_{aijk} F_{bijk} - \tf{1}{144} \eta_{ab} F_{ijkl}F_{ijkl}, \notag
  \\
  \gm^{ab} \psi_{ab} &= - \tf{1}{9} \gm_a E(\psi)^a, 
  \\
  \gm^{b} \psi_{ab} &= \tf{4}{9} E(\psi)_a - \tf{1}{18} \gm_{ab} E(\psi)^b, \notag
  \\
  D_d F^{dabc} &= 6 E(A)^{abc} + \tf{1}{24 \cdot 144} \ep_{11}^{abcijklmnop} F_{ijkl} F_{mnop}, 
  \notag
\end{alignat}
it is always possible remove terms from the ansatz which partially include
$R$, $R_{ab}$, $\gm^{ab} \psi_{ab}$, $\gm^{b} \psi_{ab}$ or $D_d F^{dabc}$.
Then the ansatz without 4-form field strength is classified into 4 kinds, 
$B_1=\mc{L}[e\hat{R}^4]_7$, 
$B_{11}=\mc{L}[e\epsilon_{11}AR^4]_{2}$,
$F_1=\mc{L}[eR^3\bar{\psi}\psi_{(2)}]_{92}$ and 
$F_2=\mc{L}[e R^2 \bar{\psi}_{(2)} D \psi_{(2)}]_{25}$.
Here $\mc{L}[X]_n$ represents a set of terms in the ansatz which become $X$ by neglecting
indices and gamma matrices. The subscript $n$ is a number of independent terms.
The variations of these terms under local supersymmetry are categorized into 3 types, 
$V_1=[eR^4\bar{\epsilon}\psi]_{116}$, 
$V_2=[eR^2DR\bar{\epsilon}\psi_{(2)}]_{88}$ and 
$V_3=[eR^3\bar{\epsilon}D\psi_{(2)}]_{51}$. 
In order to obtain these independent numbers, we have to take account of the relations
which are derived by the Bianchi identities,
\ba
  \mc{D}_{[a} \psi_{bc]} &=
  \tf{1}{4} R_{ij[ab} \gm^{ij} \psi_{c]} + DF_{[ab} \psi_{c]} + F^2_{[ab} \psi_{c]} , \label{feq}
  \\
  \gm^c \mc{D}_{c} \psi_{ab} &=
  \tf{1}{4} R_{xyab} \gm^c \gm^{xy} \psi_{c} - 2 \gm^c F_{[a} \psi_{b]c}
  + 3 \gm^c DF_{[ab} \psi_{c]} + 3 \gm^c F^2_{[ab} \psi_{c]} 
  + R_{x[a} \gm^x \psi_{b]} - 2 D_{[a} \gm^c \psi_{b]c}. \notag
\end{alignat}
These equations relate terms of $[eR^3\bar{\epsilon}D\psi_{(2)}]$ with
those of $[eR^4\bar{\epsilon}\psi]$.
Since the number of terms is more than two
hundreds, it is necessary to build a computer 
program\footnote{In refs.~\cite{HO1,HO2}, $\mc{L}[e\hat{R}^4]$ consists of 13 terms. In this paper
we put more restrictions on the form of the torsion part, and reduce the number of terms.
The equation for $\mc{D}_{b} \psi_{ab}$
is also used in ref.~\cite{HO2}. In this paper we do not use that equation since it will
make the supersymmetric transformation of the Majorana gravitino complicate.
The necessity of it might be argued when we consider the cancellation of 
the order $\mc{O}(F^2)$}.
The mechanism of the cancellation is summarized as
\begin{alignat}{5}
  &\delta \mc{L}[e\hat{R}^4]_7 &
  &\sim [eR^4\bar{\epsilon}\psi]_{116} \oplus &
  &[eR^2DR\bar{\epsilon}\psi_{(2)}]_{88}, \notag
  \\
  &\delta \mc{L}[e\epsilon_{11}AR^4]_{2} &
  &\sim [eR^4\bar{\epsilon}\psi]_{116}, \notag
  \\
  &\delta \mc{L}[eR^3\bar{\psi}\psi_{(2)}]_{92} &
  &\sim [eR^4\bar{\epsilon}\psi]_{116} \oplus &
  &[eR^2DR\bar{\epsilon}\psi_{(2)}]_{88} \oplus &
  &[eR^3\bar{\epsilon}D\psi_{(2)}]_{51}, \label{eq:var}
  \\
  &\delta \mc{L}[e R^2 \bar{\psi}_{(2)} D \psi_{(2)}]_{25} &
  &\sim &&[eR^2DR\bar{\epsilon}\psi_{(2)}]_{88} \oplus &
  &[eR^3\bar{\epsilon}D\psi_{(2)}]_{51}. \notag
\end{alignat}
Note that the variations which depend on the 4-form field strength are neglected.
The combination of the ansatz can be determined by requiring the cancellation 
of the above variations. The result is that the bosonic terms of the higher derivative
corrections are governed by two parameters as
\begin{alignat}{3}
  2 \kappa_{11}^2 S_{R^4} &= a \; \ell_p^6 \int d^{11}x \; e \Big(t_8 t_8 R^4 
  + \frac{1}{4!} \ep_{11} \ep_{11} R^4 \Big) \notag
  \\
  &\quad\,
  + b \; \ell_p^6 \int d^{11}x \; e \Big(t_8 t_8 R^4 - \frac{1}{4!} \ep_{11} \ep_{11} R^4 
  - \frac{1}{6} \ep_{11} t_8 A R^4 \Big).
\end{alignat}
The $t_8$ is a tensor with eight indices,
and $\ep_{11}$ is an antisymmetric tensor with eleven indices.
Explicit expression of the above equation can be found in ref. \cite{HO1}.
The parameters $a$ and $b$ can be determined by employing the result of four graviton
amplitude in type IIA superstring theory. In the type IIA language, non-zero $a$ corresponds
to the tree level amplitude and non-zero $b$ does to the one loop\footnote{There should be
a dilaton factor $e^{-2\phi}$ in the tree level case.}.

\section{Ansatz for $R^3F^2$ Terms and Their Variations}

In the previous section we have neglected the cancellation of terms which depend on 
the 4-form field strength $F$. Of course the cancellation of these terms
should be checked and we will execute this procedure in order.
In this paper we consider the cancellation of the variations which are linearly dependent on
the 4-form field strength. Since the variations are linear in $F$ at most, 
the ansatz for the corrections should be taken into account to the order of $F^2$. 
The terms which linearly depend on $F$ are already listed in the ansatz.
New bosonic terms which are mass dimension eight and quadratic in $F$ should be 
$B_{21}=\mc{L}[eR^3F^2]$. 
Besides this ansatz, we should also examine terms which contain more covariant derivatives,
such as $[eR^2DF^2]$, $[eDR^2F^2]$ or $[eRDRDFF]$. These 
will be important to investigate the cancellation of terms which include $DF$, 
but not so important in this discussion.

Now we added $B_{21}$ to the ansatz, it is necessary to classify independent terms of this type. 
Due to the properties of the Riemann tensor, such as the cyclicity,
it is laborious to write down all possible terms by hand. So basically we rely on a computer 
program to execute this task. 

First let us pick up a term $R_{abcd}R_{abef}R_{cdef}F_{ijkl}F_{ijkl}$ as an example.
By taking indices of this term, a list $l = \{\{a,b,c,d\},\{a,b,e,f\},
\{c,d,e,f\},\{i,j,k,l\},\{i,j,k,l\}\}$ can be made,
whose $m$-th part is denoted as $l[[m]]$.
Then we construct a 5 by 5 symmetric matrix,
\ba
  M = 
  \begin{pmatrix}
    0 & 2 & 2 & 0 & 0 \cr 2 & 0 & 2 & 0 & 0 \cr 2 & 2 & 0 & 0 & 0 \cr 
    0 & 0 & 0 & 0 & 4 \cr 0 & 0 & 0 & 4 & 0 \cr 
  \end{pmatrix},
\end{alignat}
whose $(m,n)$ component is given by the 
number of overlapping indices between $l[[m]]$ and $l[[n]]$. 
Inversely if the above matrix is given, since the indices $\{a,b,c,d\}$ of $F$ 
are uniquely placed as $F_{abcd}$ and those of $R$ are placed in two ways as
$R_{abcd}$ or $R_{acbd}$, corresponding terms of $B_{21}$ can be listed like
\begin{alignat}{3}
  &R_{abcd}R_{abef}R_{cdef}F_{ijkl}F_{ijkl}, \qquad
  R_{acbd}R_{aebf}R_{cedf}F_{ijkl}F_{ijkl}, \notag
  \\
  &R_{acbd}R_{abef}R_{cdef}F_{ijkl}F_{ijkl}, \qquad
  R_{abcd}R_{aebf}R_{cedf}F_{ijkl}F_{ijkl}, \notag
  \\
  &R_{abcd}R_{aebf}R_{cdef}F_{ijkl}F_{ijkl}, \qquad
  R_{acbd}R_{abef}R_{cedf}F_{ijkl}F_{ijkl}, 
  \\
  &R_{abcd}R_{abef}R_{cedf}F_{ijkl}F_{ijkl}, \qquad
  R_{acbd}R_{aebf}R_{cdef}F_{ijkl}F_{ijkl}. \notag
\end{alignat}
The last 6 terms are transformed into the first term up to numerical factors by applying
the relation $R_{cabd}R_{abcd}=-\frac{1}{2}R_{abcd}R_{abcd}$ repeatedly.
Therefore there are two independent terms which come from the above matrix.
In general the matrix $M$ is unique up to permutation among
three Riemann tensors or exchange of two 4-form field strengths. 
Furthermore each component of $M$ is not affected by the properties of the
Riemann tensor and the 4-form field strength. Thus in order to obtain
independent terms in $B_{21}$, it is useful to classify possible matrices
of $M$ in the first place.

Let us consider the classification of the matrices $M$.
The diagonal matrix elements are zero because the terms which contain the Ricci tensor
are excluded out of the ansatz. Each off-diagonal element is a nonnegative integer
and the maximum value is four.
The sum of the components in each row should also be four, and we obtain relations of
\begin{alignat}{3}
  &M_{12} = 4 - M_{23} - M_{24} - M_{25}, \qquad 
  M_{13} = 4 - M_{23} - M_{34} - M_{35}, \notag
  \\
  &M_{14} = 4 - M_{24} - M_{34} - M_{45}, \qquad 
  M_{15} = 4 - M_{25} - M_{35} - M_{45}, 
  \\
  &M_{23} = 6 - M_{24} - M_{25} - M_{34} - M_{35} - M_{45}. \notag
\end{alignat}
There remain five parameters of $M_{24}$, $M_{25}$, $M_{34}$, $M_{35}$ and $M_{45}$.
Note, however, that the range of
former 4 parameters should be from $0$ to $2$, because $R_{abcd}F_{abce} = 0$.
Then, up to the permutation among three Riemann tensors or the exchange of two 4-form 
field strengths, there are 18 possible matrices of $M$,
\ba
  &M[1]=
  \begin{pmatrix}
    0 & 0 & 0 & 2 & 2 \cr 0 & 0 & 2 & 0 & 2 \cr 0 & 2 & 0 & 2 & 0 \cr 
    2 & 0 & 2 & 0 & 0 \cr 2 & 2 & 0 & 0 & 0 \cr 
  \end{pmatrix}, \;
  M[2]=
  \begin{pmatrix}
    0 & 0 & 0 & 2 & 2 \cr 0 & 0 & 2 & 1 & 1 \cr 0 & 2 & 0 & 1 & 1 \cr 
    2 & 1 & 1 & 0 & 0 \cr 2 & 1 & 1 & 0 & 0 \cr 
  \end{pmatrix}, \;
  M[3]=
  \begin{pmatrix}
    0 & 0 & 0 & 2 & 2 \cr 0 & 0 & 3 & 0 & 1 \cr 0 & 3 & 0 & 1 & 0 \cr 
    2 & 0 & 1 & 0 & 1 \cr 2 & 1 & 0 & 1 & 0 \cr 
  \end{pmatrix}, \notag
  \\
  &M[4]=
  \begin{pmatrix}
    0 & 0 & 0 & 2 & 2 \cr 0 & 0 & 4 & 0 & 0 \cr 0 & 4 & 0 & 0 & 0 \cr 
    2 & 0 & 0 & 0 & 2 \cr 2 & 0 & 0 & 2 & 0 \cr 
  \end{pmatrix}, \;
  M[5]=
  \begin{pmatrix}
    0 & 0 & 1 & 1 & 2 \cr 0 & 0 & 1 & 1 & 2 \cr 1 & 1 & 0 & 2 & 0 \cr 
    1 & 1 & 2 & 0 & 0 \cr 2 & 2 & 0 & 0 & 0 \cr 
  \end{pmatrix}, \;
  M[6]=
  \begin{pmatrix}
    0 & 0 & 1 & 1 & 2 \cr 0 & 0 & 1 & 2 & 1 \cr 1 & 1 & 0 & 1 & 1 \cr 
    1 & 2 & 1 & 0 & 0 \cr 2 & 1 & 1 & 0 & 0 \cr 
  \end{pmatrix}, \notag
  \\
  &M[7]=
  \begin{pmatrix}
    0 & 0 & 1 & 1 & 2 \cr 0 & 0 & 2 & 1 & 1 \cr 1 & 2 & 0 & 1 & 0 \cr 
    1 & 1 & 1 & 0 & 1 \cr 2 & 1 & 0 & 1 & 0 \cr 
  \end{pmatrix}, \;
  M[8]=
  \begin{pmatrix}
    0 & 0 & 1 & 1 & 2 \cr 0 & 0 & 2 & 2 & 0 \cr 1 & 2 & 0 & 0 & 1 \cr 
    1 & 2 & 0 & 0 & 1 \cr 2 & 0 & 1 & 1 & 0 \cr 
  \end{pmatrix}, \;
  M[9]=
  \begin{pmatrix}
    0 & 0 & 1 & 1 & 2 \cr 0 & 0 & 3 & 1 & 0 \cr 1 & 3 & 0 & 0 & 0 \cr 
    1 & 1 & 0 & 0 & 2 \cr 2 & 0 & 0 & 2 & 0 \cr 
  \end{pmatrix}, 
  \\
  &M[10]=
  \begin{pmatrix}
    0 & 0 & 2 & 0 & 2 \cr 0 & 0 & 2 & 2 & 0 \cr 2 & 2 & 0 & 0 & 0 \cr 
    0 & 2 & 0 & 0 & 2 \cr 2 & 0 & 0 & 2 & 0 \cr 
  \end{pmatrix}, \;
  M[11]=
  \begin{pmatrix}
    0 & 0 & 2 & 1 & 1 \cr 0 & 0 & 2 & 1 & 1 \cr 2 & 2 & 0 & 0 & 0 \cr 
    1 & 1 & 0 & 0 & 2 \cr 1 & 1 & 0 & 2 & 0 \cr 
  \end{pmatrix}, \;
  M[12]=
  \begin{pmatrix}
    0 & 1 & 1 & 0 & 2 \cr 1 & 0 & 1 & 1 & 1 \cr 1 & 1 & 0 & 2 & 0 \cr 
    0 & 1 & 2 & 0 & 1 \cr 2 & 1 & 0 & 1 & 0 \cr 
  \end{pmatrix}, \notag
  \\
  &M[13]=
  \begin{pmatrix}
    0 & 1 & 1 & 0 & 2 \cr 1 & 0 & 2 & 1 & 0 \cr 1 & 2 & 0 & 1 & 0 \cr 
    0 & 1 & 1 & 0 & 2 \cr 2 & 0 & 0 & 2 & 0 \cr 
  \end{pmatrix}, \;
  M[14]=
  \begin{pmatrix}
    0 & 1 & 1 & 1 & 1 \cr 1 & 0 & 1 & 1 & 1 \cr 1 & 1 & 0 & 1 & 1 \cr 
    1 & 1 & 1 & 0 & 1 \cr 1 & 1 & 1 & 1 & 0 \cr 
  \end{pmatrix}, \;
  M[15]=
  \begin{pmatrix}
    0 & 1 & 1 & 1 & 1 \cr 1 & 0 & 2 & 0 & 1 \cr 1 & 2 & 0 & 1 & 0 \cr 
    1 & 0 & 1 & 0 & 2 \cr 1 & 1 & 0 & 2 & 0 \cr 
  \end{pmatrix}, \notag
\end{alignat}
\ba
  &M[16]=
  \begin{pmatrix}
    0 & 1 & 1 & 1 & 1 \cr 1 & 0 & 3 & 0 & 0 \cr 1 & 3 & 0 & 0 & 0 \cr 
    1 & 0 & 0 & 0 & 3 \cr 1 & 0 & 0 & 3 & 0 \cr 
  \end{pmatrix}, \;
  M[17]=
  \begin{pmatrix}
    0 & 1 & 2 & 0 & 1 \cr 1 & 0 & 2 & 1 & 0 \cr 2 & 2 & 0 & 0 & 0 \cr 
    0 & 1 & 0 & 0 & 3 \cr 1 & 0 & 0 & 3 & 0 \cr 
  \end{pmatrix}, \;
  M[18]=
  \begin{pmatrix}
    0 & 2 & 2 & 0 & 0 \cr 2 & 0 & 2 & 0 & 0 \cr 2 & 2 & 0 & 0 & 0 \cr 
    0 & 0 & 0 & 0 & 4 \cr 0 & 0 & 0 & 4 & 0 \cr 
  \end{pmatrix}. \notag
\end{alignat}
The result is generated by using a computer program, though it is not so 
difficult to derive it by hand. 

We are now ready to classify terms in $B_{21}$.
As explained in the example, eight terms are assigned for each matrix of the above.
By using the properties of the Riemann tensor, some of these terms will be transformed 
into some combination of the other terms. This manipulation is systematic and 
is executed again by employing the computer program. 
In this way we can write down independent terms for each matrix, 
and finally obtain 30 terms for $B_{21}$,
\begin{alignat}{3}
  &B_{21}[1] = R_{m n i j} R_{a b o p} R_{a b k l} F_{i j k l} F_{m n o p}, \qquad&
  &B_{21}[2] = R_{m n i j} R_{a b k o} R_{a b l p} F_{i j k l} F_{m n o p}, \notag
  \\
  &B_{21}[3] = R_{m n i j} R_{a k b o} R_{a l b p} F_{i j k l} F_{m n o p}, \qquad&
  &B_{21}[4] = R_{l m i j} R_{a b c n} R_{a b c k} F_{w i j k} F_{w l m n}, \notag
  \\
  &B_{21}[5] = R_{k l i j} R_{a b c d} R_{a b c d} F_{w x i j} F_{w x k l}, \qquad&
  &B_{21}[6] = R_{i a m n} R_{j b o p} R_{a b k l} F_{i j k l} F_{m n o p}, \notag
  \\
  &B_{21}[7] = R_{i a m n} R_{o b j k} R_{a b l p} F_{i j k l} F_{m n o p}, \qquad&
  &B_{21}[8] = R_{i a m n} R_{o b j k} R_{a l b p} F_{i j k l} F_{m n o p}, \notag
  \\
  &B_{21}[9] = R_{i a l m} R_{b c j n} R_{a b c k} F_{w i j k} F_{w l m n}, \qquad&
  &B_{21}[10] = R_{i a l m} R_{b j c n} R_{a b c k} F_{w i j k} F_{w l m n}, \notag
  \\
  &B_{21}[11] = R_{i a l m} R_{b c j k} R_{a b c n} F_{w i j k} F_{w l m n}, \qquad&
  &B_{21}[12] = R_{i a k l} R_{b c d j} R_{a b c d} F_{w x i j} F_{w x k l}, \notag
  \\
  &B_{21}[13] = R_{a b k l} R_{c d i j} R_{a b c d} F_{w x i j} F_{w x k l}, \qquad&
  &B_{21}[14] = R_{a b i k} R_{c d j l} R_{a b c d} F_{w x i j} F_{w x k l}, \notag
  \\
  &B_{21}[15] = R_{a i b k} R_{c j d l} R_{a c b d} F_{w x i j} F_{w x k l}, \qquad&
  &B_{21}[16] = R_{a b l m} R_{a c i n} R_{b c j k} F_{w i j k} F_{w l m n}, 
  \\
  &B_{21}[17] = R_{a b l m} R_{a i c n} R_{b c j k} F_{w i j k} F_{w l m n}, \qquad&
  &B_{21}[18] = R_{a b k l} R_{a c d i} R_{b c d j} F_{w x i j} F_{w x k l}, \notag
  \\
  &B_{21}[19] = R_{a b k l} R_{a c d i} R_{b d c j} F_{w x i j} F_{w x k l}, \qquad&
  &B_{21}[20] = R_{a b i l} R_{a c j m} R_{b k c n} F_{w i j k} F_{w l m n}, \notag
  \\
  &B_{21}[21] = R_{a i b l} R_{a j c m} R_{b k c n} F_{w i j k} F_{w l m n}, \qquad&
  &B_{21}[22] = R_{a b i k} R_{a c d l} R_{b c d j} F_{w x i j} F_{w x k l}, \notag
  \\
  &B_{21}[23] = R_{a b i k} R_{a c d l} R_{b d c j} F_{w x i j} F_{w x k l}, \qquad&
  &B_{21}[24] = R_{a i b k} R_{a c d l} R_{b c d j} F_{w x i j} F_{w x k l}, \notag
  \\
  &B_{21}[25] = R_{a i b k} R_{a c d l} R_{b d c j} F_{w x i j} F_{w x k l}, \qquad&
  &B_{21}[26] = R_{a i b j} R_{a c d e} R_{b c d e} F_{w x y i} F_{w x y j}, \notag
  \\
  &B_{21}[27] = R_{a b c j} R_{a d e i} R_{b c d e} F_{w x y i} F_{w x y j}, \qquad&
  &B_{21}[28] = R_{a b c j} R_{a d e i} R_{b d c e} F_{w x y i} F_{w x y j}, \notag
  \\
  &B_{21}[29] = R_{a b c d} R_{a b e f} R_{c d e f} F_{w x y z} F_{w x y z}, \qquad&
  &B_{21}[30] = R_{a c b d} R_{a e b f} R_{c e d f} F_{w x y z} F_{w x y z}. \notag
\end{alignat}
The matrix $M[15]$ generates four terms of $B_{21}[22]$, $B_{21}[23]$, $B_{21}[24]$ 
and $B_{21}[25]$, and other matrices generate two terms at most.

The variations of the terms in $B_{21}$ can be obtained by applying the 
supersymmetric transformations (\ref{eq:susytr}).
Since in this paper we are concerned with the cancellation of the variations which are 
linear in $F$, we only vary the 4-form field strength.
The variations which include $DF$ are also neglected. Then the variations of the terms in $B_{21}$
under the local supersymmetry are sketched as
\begin{alignat}{3}
  \delta \mc{L}[eR^3F^2]_{30} &\sim [eR^2DRF\bar{\epsilon}\psi].
\end{alignat}
For example, the variation of $B_{21}[30]$ is calculated as
\ba
  \dl (e R_{a c b d} R_{a e b f} R_{c e d f} F_{w x y z} F_{w x y z}) \sim 
  72 \, e R_{a c b d} R_{a e b f} D_w R_{c e d f} F_{w x y z} \bar{\ep} \gm_{xy} \psi_z.
\end{alignat}
Here the partial integral is used so that the covariant derivative does not act
on the fermionic parameter. The variations of remaining 29 terms can also be obtained by hand.
Due to the properties of the Riemann tensor, however, those expressions are complicatedly
related to each other in general. Therefore it is necessary to classify
all independent terms in $V_{11}=[eR^2DRF\bar{\epsilon}\psi]_{1563}$.
Compared to the bosonic case, the classification of this type is much more laborious
because of the existence of additional indices coming from gamma matrices.
Then it in inevitable to employ a computer program, and the output shows that
$V_{11}$ contains 1563 independent terms.

\section{Result of the Cancellation}

Our concern is the cancellation of the variations to the order of $F$.
The ansatz considered so far is $B_{1}$, $B_{11}$, $F_{1}$, $F_{2}$ and $B_{21}$.
The variations of $B_{21}$ make the terms in $V_{11}$ which are linear in the 4-form field
strength. The variations of $B_{1}$, $B_{11}$, $F_{1}$ and $F_{2}$ consist of the terms in
$V_{11}$ and $V_{12}=[eR^3F\bar{\ep}\psi_{(2)}]_{513}$.
The classification of $V_{12}$ is also executed by using the computer program,
and the output shows that there are 513 independent terms.

Unfortunately the variations of the above ansatz do not cancel completely, so we need to
add more terms to the ansatz. These terms are bilinear in the Majorana gravitino and
listed as
\begin{alignat}{3}
  &F_{11}=\mc{L}[eR^3F\bar{\psi}\psi]_{447}, \qquad&
  &F_{12}=\mc{L}[eR^2F\bar{\psi}_{(2)}\psi_{(2)}]_{190}, \notag
  \\
  &F_{13}=\mc{L}[eR^2DF\bar{\psi}\psi_{(2)}]_{614}, \qquad&
  &F_{14}=\mc{L}[eRDF\bar{\psi}_{(2)}D\psi_{(2)}]_{113}.
\end{alignat}
The subscript number represents that of independent terms.
The variations of these terms consist of $V_{11}$, $V_{12}$ and 
$V_{16}=[eR^2DDF\bar{\ep}\psi_{(2)}]_{151}$. The last type $V_{16}$ is taken into account
since these terms are potentially related to $V_{12}$ by the relation (\ref{eq:comm}).
Detailed explanation will be given in ref. \cite{H1}.
The cancellation mechanism is sketched in the following table.
\begin{alignat}{5}
  &\delta \mc{L}[eR^4]_7 &
  &\sim [eR^2DRF\bar{\epsilon}\psi]_{1563} , \notag
  \\
  &\delta \mc{L}[e\epsilon_{11}AR^4]_{2} &
  &\sim [eR^2DRF\bar{\epsilon}\psi]_{1563}, \notag
  \\
  &\delta \mc{L}[eR^3\bar{\psi}\psi_{(2)}]_{92} &
  &\sim &
  &[eR^3F\bar{\epsilon}\psi_{(2)}]_{513} , \notag
  \\
  &\delta \mc{L}[e R^2 \bar{\psi}_{(2)} D \psi_{(2)}]_{25} &
  &\sim &&[eR^3F\bar{\epsilon}\psi_{(2)}]_{513}, \notag
  \\
  &\delta \mc{L}[eR^3F^2]_{30} &
  &\sim [eR^2DRF\bar{\epsilon}\psi]_{1563} , \label{eq:var}
  \\
  &\delta \mc{L}[eR^3F\bar{\psi}\psi]_{447} &
  &\sim [eR^2DRF\bar{\epsilon}\psi]_{1563} \oplus &
  &[eR^3F\bar{\epsilon}\psi_{(2)}]_{513} , \notag
  \\
  &\delta \mc{L}[e R^2 F \bar{\psi}_{(2)} \psi_{(2)}]_{190} &
  &\sim &&[eR^3F\bar{\epsilon}\psi_{(2)}]_{513}, \notag
  \\
  &\delta [e R^2 DF \bar{\psi} \psi_{(2)}]_{614} &
  &\sim &&[eR^3F\bar{\epsilon}\psi_{(2)}]_{513} \oplus &
  &[eR^2DDF\bar{\epsilon}\psi_{(2)}]_{151}, \notag
  \\
  &\delta \mc{L}[e R DF \bar{\psi}_{(2)} D \psi_{(2)}]_{113} &
  &\sim &&[eR^3F\bar{\epsilon}\psi_{(2)}]_{513} \oplus &
  &[eR^2DDF\bar{\epsilon}\psi_{(2)}]_{151}. \notag
\end{alignat}
Note that we only considered the variations which are linear in $F$.
If we examine the cancellation of the terms which include $DF$, 
we need to add $\mc{L}[R^2DF^2]$ to the ansatz\footnote{The variations of 
$\mc{L}[R^2DF^2]$ will also include $V_{16}$. These terms are neglected in 
this paper but will be considered in the other place.}.

The cancellation among the variations of $V_1$, $V_2$, $V_3$, $V_{11}$, $V_{12}$ and 
$V_{16}$ is examined by building a computer program. 
There are 2482 linear equations among 1520 coefficients in the ansatz, and
solution for the bosonic part is written as
\begin{alignat}{3}
  &2 \kappa_{11}^2 S_{R^4} \notag
  \\
  &= \ell_p^6 \int d^{11}x \; e \bigg\{
  \frac{b}{24} \Big(t_8 t_8 R^4 - \frac{1}{4!} \ep_{11} \ep_{11} R^4 
  - \frac{1}{6} \ep_{11} t_8 A R^4 \Big) \notag
  \\
  &\quad\,
  + \Big( -\frac{b_{6}}{4} + \frac{b_{10}}{16} - \frac{b_{11}}{8} - b_{13} 
  - \frac{b_{22}}{8} - \frac{b_{23}}{8} - \frac{b_{24}}{8} - \frac{b_{25}}{8} 
  + \frac{3 b_{28}}{8} + \frac{9 b_{30}}{2} \Big) B_{21}[1] \notag
  \\
  &\quad\,
  +\Big( - \frac{68 b}{3} - \frac{b_{6}}{2} - \frac{b_{7}}{2} - \frac{b_{9}}{4} 
  - \frac{3 b_{10}}{8} - \frac{b_{11}}{4} + \frac{b_{17}}{2} + \frac{b_{19}}{2} -
  \frac{b_{21}}{8} + \frac{b_{22}}{2} + \frac{b_{23}}{2} + \frac{b_{24}}{2} \notag
  \\
  &\quad\,
  + \frac{b_{25}}{4} - \frac{3 b_{28}}{2} - 30 b_{30} \Big) B_{21}[2] 
  + \Big( \frac{68 b}{3} + b_{7} + b_{8} + \frac{b_{10}}{2} - b_{16} - b_{17} - b_{18} 
  - b_{19} - \frac{b_{21}}{4} \notag
  \\
  &\quad\,
  - \frac{b_{22}}{2} - \frac{b_{23}}{2} - \frac{b_{24}}{2} 
  - \frac{b_{25}}{2} + \frac{3 b_{28}}{2} + 30 b_{30} \Big) B_{21}[3] 
  + \Big( b_{19} - 6 b_{26} - 36 b_{30} \Big) B_{21}[12] \notag
  \\
  &\quad\,
  + \Big( \frac{b_{22}}{2} + \frac{b_{23}}{4} + \frac{b_{24}}{4} -
  \frac{3 b_{27}}{4} - \frac{3 b_{28}}{2} \Big) B_{21}[14] 
  + \Big(-b_{22} - b_{23} - b_{24} - b_{25} + 3 b_{28} \label{eq:result}
  \\
  &\quad\,
  + 36 b_{30} \Big) B_{21}[15] 
  + \Big( -\frac{32 b}{3} + 2 b_{22} + 2 b_{23} - 6 b_{28} - 48 b_{30} \Big) B_{21}[20] 
  - \frac{b_{30}}{2} B_{21}[29] \notag
  \\
  &\quad\,
  + b_{4} B_{21}[4] + b_{5} B_{21}[5] + b_{6} B_{21}[6] + b_{7} B_{21}[7] 
  + b_{8} B_{21}[8] + b_{9} B_{21}[9] + b_{10} B_{21}[10] \notag
  \\
  &\quad\,
  + b_{11} B_{21}[11] + b_{13} B_{21}[13] + b_{16} B_{21}[16] 
  + b_{17} B_{21}[17] + b_{18} B_{21}[18] + b_{19} B_{21}[19] \notag
  \\
  &\quad\,
  + b_{21} B_{21}[21] + b_{22} B_{21}[22] + b_{23} B_{21}[23] 
  + b_{24} B_{21}[24] + b_{25} B_{21}[25] + b_{26} B_{21}[26] \notag
  \\
  &\quad\,
  + b_{27} B_{21}[27] + b_{28} B_{21}[28] 
  + b_{30} B_{21}[30] \bigg\}. \notag
\end{alignat}
Remarkable result is that the combination of the $R^4$ terms is completely
fixed by the local supersymmetry. The structure of these terms is consistent 
with one-loop calculations in the type IIA superstring theory. 
On the other hand the structure of the $R^3F^2$ terms 
is not so clear. In order to determine a combination of these terms, we need to
investigate the cancellation of the terms which include $DF$ or $F^2$.

\section{Conclusion and Discussion}

The higher derivative terms in the M-theory are investigated by applying
the Noether method. The cancellation of the variations under local supersymmetry
is examined to the order linear in the 4-form field strength $F$. 
Since the calculations are hard, we heavily employed the computer
program to check the cancellation.

The bosonic part of the ansatz consists of 39 terms, $B_{1}$, $B_{11}$ and $B_{21}$, and 
the fermionic part of the ansatz does of 1481 terms, $F_{1}$, $F_{2}$, $F_{11}$,
$F_{12}$, $F_{13}$ and $F_{14}$. The variations of the ansatz are expanded by 2482 terms,
$V_{1}$, $V_{2}$, $V_{3}$, $V_{11}$, $V_{12}$ and $V_{16}$.
By requiring the cancellation of these variations, we obtain 2482 linear equations among
1520 coefficients in the ansatz.
Then the coefficients of the bosonic part is solved as the eq.~(\ref{eq:result}).
Remarkably the structure of the $R^4$ terms is uniquely determined by the requirement 
of the local supersymmetry. This result exactly matches with the fact that the one-loop effective 
action in the type IIA superstring theory survives after taking $g_s \to \infty$.

On the other hand $R^3F^2$ terms cannot be fixed completely. 
In order to determine the structure of these terms, we need to proceed to the cancellation 
of terms which depend on $DF$ or $F^2$. The cancellation of the terms which contain $DF$
will require $R^2DF^2$ terms in the ansatz, and it may fix both $R^3F^2$ terms and $R^2DF^2$ terms.
Then it becomes possible to compare the result with 
that obtained by the scattering amplitudes of type IIA superstring theory\cite{PVW2}.
It is also important to check the consistency to the result obtained by 
the superspace formalism or the IIB matrix model\cite{Ra,KMS}.

As a conclusion the local supersymmetry seems to determine the structure of the higher derivative
corrections in M-theory uniquely. Similar statement can be found in the context of 
D-particle dynamics\cite{KM}. We will succeed the procedure executed in this paper and determine 
the structure of the higher derivative corrections in M-theory completely. 
After the determination of the action, applications to black hole 
physics or cosmology will become interesting future directions\cite{My,MO}.

\section*{Acknowledgements}

I would like to thank Sachiko Ogushi for the collaboration and encouragement, and
thank members of particle theory group in Osaka University,
especially K. Hotta, T. Kubota, T. Nakatsu, Y. Noma, M. Sato and A. Tsuchiya
for discussions or comments.
The mathematica code ``GAMMA.m'' is useful to calculate products of gamma matrices\cite{Gr}.
I also thank the Yukawa Institute for Theoretical Physics at Kyoto University. 
Discussions during the YITP workshop YITP-W-06-11 on 
"String Theory and Quantum Field Theory" were useful to complete this work.
Discussions during the KEK theory workshop were also useful for future directions of this work,
and I would like to thank L. Dixon, M. Hamanaka, K. Hotta, H. Kunitomo, Y. Michishita, N. Ohta, 
T. Sakai, S. Terashima and S. Yahikozawa.
This work is supported in part by The 21st Century COE Program 
"Towards a New Basic Science; Depth and Synthesis.''

\end{document}